\documentstyle[12pt,epsfig]{article}
\input epsf
\begin{document}
\title{Mechanisms of the $f_0(980)$ production in the reaction $\pi^-p\to\pi^0
\pi^0n$\\[1cm]}
\author{N.N. Achasov \thanks{Email address: achasov@math.nsc.ru}\,\,
and G.N. Shestakov \thanks{Email address: shestako@math.nsc.ru}\\[0.8cm]
{\it Laboratory of Theoretical Physics,}\\
{\it S.L. Sobolev Institute for Mathematics,}\\
{\it 630090, Novosibirsk 90,Russia}\\[1cm]}
\date{}
\maketitle\begin{abstract}
The model of the pure one-pion exchange mechanism, which gives a good 
description of the GAMS results on the alteration of the $S$-wave $\pi^0\pi^0$ 
mass spectrum in the $f_0(980)$ region in the reaction $\pi^-p\to\pi^0\pi^0n$
with increasing $-t$, is compared with the recent detailed data on the $m$ and
$t$ distributions of the $\pi^-p\to\pi^0\pi^0n$ events obtained by the BNL-E852
Collaboration. It is shown that the predictions of this model are not confirmed
by the BNL data. Therefore the observed phenomenon should be explained by the
different exchange mechanism. It is most likely to be the $a_1$ exchange
mechanism.
\end{abstract} \vspace{0.8cm}
\hspace{1.1cm}PACS number(s): 12.40.Nn, 13.75.Lb, 13.85.Hd
\newpage
\begin{center}{\bf I. INTRODUCTION}\end{center}

In two recent experiments on the reaction $\pi^-p\to\pi^0\pi^0n$ at high
energies performed by the GAMS Collaboration at Institute of High Energy
Physics [1,2] and the BNL-E852 Collaboration at Brookhaven National Laboratory
(BNL) [3,4] it has been found a very interesting phenomenon consisting in the
alteration of the $S$-wave $\pi^0\pi^0$ mass spectrum in the vicinity of the
$f_0(980)$ resonance with increasing of $-t$, where $t$ is the square of the
four-momentum transferred from the incoming $\pi^-$ to the outgoing $\pi^0\pi^0
$ system. If, for small values of $-t$, where the reaction $\pi^-p\to\pi^0\pi^0
n$ is dominated by the one-pion exchange mechanism, the $f_0(980)$
resonance manifests itself in the $S$-wave $\pi^0\pi^0$ mass spectrum as a dip
due to its strong destructive interference with the large and smooth
background, then, for large values of $-t$, it appears as a peak [1-4].

The GAMS and BNL-E852 results are based on high statistics and impose severe 
demands on the phenomenological models constructed for their explanation.

Historically, the first description of the GAMS results on the $f_0(980)$
resonance [1] has been performed in Ref. [5] on the basis of the pure one-pion
exchange (POPE) model. To explain the observed dip and peak behavior of the $f
_0(980)$ in this model the authors of Ref. [5] had to provide the individual
contributions to the full $S$-wave $\pi^{*+}(t)\pi^-\to\pi^0\pi^0$ amplitude
(where $\pi^*(t)$ is a Reggeized pion) with rather exotic $t$ dependencies. 
With minor modifications such a treatment of the GAMS data [1] has been also
reproduced in a series of the subsequent publications [6,7]. Then, we suggested
a crucially new explanation of the GAMS results on the $f_0(980)$ production in
which a main role was assigned to the $\pi^-p\to f_0(980)n$ reaction amplitude
with the quantum numbers of the $a_1$ Regge pole in the $t$ channel [8]. In
brief, our scenario came to the following. At small $-t$, the reaction $\pi^-p
\to(\pi^0\pi^0)_S\,n$ [hereafter $(\pi\pi)_S$ denotes a $\pi\pi$ system with
the orbital angular momentum $L=0$] is dominated by the one-pion exchange
mechanism, and the $f_0(980)$ resonance appears as a minimum in the $(\pi^0\pi
^0)_S$ mass spectrum. However, the one-pion exchange contribution decreases
very rapidly with $-t$ and the $a_1$ exchange mechanism becomes the dominant
one in the reaction $\pi^-p\to(\pi^0\pi^0)_S\,n$ at large $-t$ [8].
The $f_0(980)$ resonance produced via the $a_1$ exchange shows
itself as a peak in the $(\pi^0\pi^0)_S$ mass spectrum because precisely such a
manifestation of the $f_0(980)$ has been observed in all known
reactions in which the $f_0(980)$ production channel differs from that of the
elastic $\pi\pi$ interaction [8].

In spite of the quite satisfactory descriptions of the GAMS data in Refs. [5]
and [8], both above mentioned models are certainly in need of further
experimental tests [8]. Note that the scenario considered in Ref. [8] can be
rejected only by the measurements of the reaction $\pi^-p\to(\pi^0\pi^0)_S\,n$
on polarized targets because only they will make it possible an explicit
separation of the $\pi$ and $a_1$ exchange mechanisms. As to the test of the
POPE model [5], it can be easily fulfilled experimentally owing to the specific
predictions of this model, for example, for the $t$ distributions of the $\pi^-
p\to(\pi^0\pi^0)_S\,n$ events in the region $0<-t<0.2$ GeV$^2$ for $m<1$ GeV 
[where $m$ is the invariant mass of the $(\pi^0\pi^0)_S$ system]. In part, we 
have already drawn attention to this circumstance in Ref. [8]. It is necessary
to note that the comparison with the available GAMS data does not allow to 
reveal all predictions hidden in the POPE model [5]. Fortunately, the 
recent data on the $m$ and $t$ distributions of the $\pi^-p\to(\pi^0\pi^0)_S\,n
$ events presented by the BNL-E852 Collaboration [4] give a unique possibility
to carry out the detailed comparison of the POPE model with experiment. Such a
comparison is the main goal of this work.

In Sec. II, we briefly recall the initial POPE model constructed in Ref. [5]
for the description of the dip and peak behavior of the $(\pi^0\pi^0)_S$ mass
spectrum in the $f_0(980)$ region. All subsequent versions [6,7,9,10] of this
model are also briefly discussed. We emphasize that the POPE model [5]
leads to a full violation of the $t$ dependence factorization hypothesis for
the $S$-wave $\pi^{*}(t)\pi\to\pi\pi$ amplitude. In contrast, this hypothesis,
as it is well known, has been widely used previously as a simple and reliable
working tool for obtaining the data on the lower $\pi\pi$ scattering partial 
waves (see, for example, Refs. [8,11-16]). Here we also discuss possibilities
of the unambiguous experimental verification of the POPE model predictions 
associated with the above violation. In Sec. III, we perform a detailed 
comparison of the POPE model [5] with the BNL data [4]. Our conclusions are 
briefly summarized in Sec. IV.

\begin{center}{\bf II. MODEL OF THE ONE-PION EXCHANGE AMPLITUDE FOR THE
REACTION {\boldmath $\pi^-p\to(\pi^0\pi^0)_S\,n$}}\end{center}

The double differential distribution in $m$ and $t$ of the $\pi^-p\to(\pi^0\pi
^0)_S\,n$ reaction events at fixed incident pion momentum is defined by the
authors of the POPE model [5] as follows \begin{eqnarray} \frac{d^2N}{
dmdt}=C\,\left|\,\frac{\sqrt{-t}}{m^2_\pi-t}\ F(t)\ a_{\pi\pi}(m,t)\,\right|^2
\,,\end{eqnarray} where $C$ is the normalization constant, $F(t)$ is the form
factor pertaining to the $\pi^*(t)NN$ vertex and $a_{\pi\pi}(m,t)$ is the $S
$-wave $\pi^*(t)\pi\to\pi\pi$ amplitude with isospin $I=0$. To construct the
amplitude $a_{\pi\pi}(m,t)$ the $K$ matrix method was used in Ref. [5], and to
describe the data in the region $0.7<m<1.2$ GeV contributions of two resonances
coupled to the $\pi\pi$ and $K\bar K$ channels and some background terms were
taken into account in the $K$ matrix. From the general formula for the
amplitude $\hat A=\hat K(t)(I-i\hat\rho\hat K)^{-1}$, where $\hat A$ and $\hat
K$ are $2\times2$ matrices describing the transitions in the $\pi\pi$ and $K
\bar K$ channels, and $\hat\rho$ is a diagonal matrix of the phase volumes, it
follows that
\begin{equation} a_{\pi\pi}(m,t)=\frac{K_{\pi\pi}(t)+i\rho_K[K_{\pi K}(t)K_{K
\pi}-K_{\pi\pi}(t)K_{K\bar K}]}{1-i\rho_\pi K_{\pi\pi}-i\rho_K K_{K\bar K}+\rho
_\pi\rho_K[K_{\pi K}K_{K\pi}-K_{\pi\pi}K_{K\bar K}]}\,,\end{equation} where,
according to Ref. [5], $\rho_\pi=(1-4m^2_\pi/m^2)^{1/2}$,\, $\rho_K=(1-4m^2_K/m
^2)^{1/2}$\, ($\rho_K\to i|\rho_K|$ for $0<m<2m_K$\,), $K_{ab}=K_{ab}(t=m^2_\pi
)\,$, $K_{\pi K}=K_{K\pi}\,$, \begin{equation} K_{ab}(t)=\left[\frac{g_a(t)\,g
_b}{M^2_1-m^2}+\frac{G_a(t)\,G_b}{M^2_2-m^2}+f_{ab}(t)\right]\left(1-\frac{m^2
_\pi}{2m^2}\right)\,,\end{equation} $f_{K\bar K}(t)=0\,$ ($a=\pi,K\,$; $\,b=\pi
,K,\bar K\,$; $\,g_{\bar K}=g_K\,$ and $\,G_{\bar K}=G_K$).
\footnote{Using the representation (2), it is easy to show that the authors of
Ref. [5] missed in Eq. (1) the $m$ dependent factor $m\rho_\pi$ which is 
approximately equal to 1 only in the vicinity of $m=1$ GeV.}

In order to simplify the discussion of the expression (2), it is convenient,
for the moment, to neglect in Eq. (3) all background terms $f_{ab}(t)$ and the
quantity $m^2_\pi/2m^2$ which is negligible for $m\approx1$
GeV. With these simplifications in mind, Eq. (2) can be rewritten in the
following more transparent form: \begin{equation} a_{\pi\pi}(m,t)=
\frac{g_\pi(t)\,[D_2(m)\,g_\pi+\Pi_{12}(m)\,G_\pi]+G_\pi(t)\,[D_1(m)\,G_
\pi+\Pi_{12}(m)\,g_\pi]}{D_1(m)D_2(m)-\Pi^2_{12}(m)}\,,\end{equation} where $D
_1(m)=M^2_1-m^2-ig_\pi^2\rho_\pi-ig_\pi^2\rho_K$ and $D_2(m)=M^2_2-m^2-iG_\pi^2
\rho_\pi-iG_K^2\rho_K$ are the inverse propagators for the initial bare
resonances, and $\Pi_{12}(m)=ig_\pi G_\pi\rho_\pi+ig_KG_K\rho_K$ is the
amplitude describing the transitions between these resonances through the real
$\pi\pi$ and $K\bar K$ intermediate states. In Eq. (4), it is easily recognized
the amplitude of the process $\pi^*(t)\pi\to\pi\pi$ with $L=I=0$ due to the
contributions of two mixed resonances coupled to the $\pi\pi$ and $K\bar K$
channels.

It is now well understood that the observed alteration of the  $(\pi^0\pi^0)_S$
mass spectrum can be described with the considered model only if the
destructive interference between two resonances at $m\approx1$ GeV, which
occurs in the low $-t$ region, is replaced by the constructive one with
increasing $-t$. According to Eq. (4), this means a change of the interference
type between the terms proportional to $g_\pi(t)$ and $G_\pi(t)$. In its turn, 
this
is possible only if, as $-t$ increases, one of the residues, for example, $g
_\pi(t)$, decreases in absolute value, vanishes at a certain value $t=t_0$, and
then changes its sign. According to the fit to the GAMS data presented in
Ref. [5], this has to occur for $-t<0.2$ GeV$^2$. Hence, due to
such an approach, the $t$ dependence of the amplitude $a_{\pi\pi}(m,t)$ must
not factorize at $m\approx1$ GeV even in the low $-t$ region. Here, in addition
to the remark mentioned in the Introduction about the $t$ dependence
factorization hypothesis, we note that the results on the $\pi\pi$ scattering
obtained by using this hypothesis were always in close agreement with those of
the more general Chew-Low extrapolation method [11-16]. In its simplest and
most frequently used form [8,11-16], the factorization hypothesis implies in
this case that, at least  for small values of $-t$, i.e. in the region $0<-t<
(0.15-0.20)$ GeV$^2$, the amplitude $a_{\pi\pi}(m,t)$ is proportional to the
on-mass-shell amplitude $a_{\pi\pi}(m,t=m^2_\pi)$. In doing so, the factor
of proportionality is generally taken in the form $\exp[b(t-m_\pi^2)]$. On
the other hand, if one explains the GAMS data in the framework of the POPE
model [5], then the factorization hypothesis must be rejected from the outset.

In Ref. [5], the following parametrization for the residues $g_\pi(t)$, $G_\pi(
t)$, $f_{\pi\pi}(t)$, and $f_{\pi K}(t)$ was postulated: \begin{equation} g_\pi
(t)=g_\pi+(1-t/m^2_\pi)\,t\,g'_\pi/m^2_\pi\ , \qquad G_\pi(t)=G_\pi+(1-t/m^2
_\pi)\,t\,G'_\pi/m^2_\pi\ ,\end{equation} \begin{equation}\ \ \ f_{\pi\pi}(t)=(
1-t/m^2_\pi)\,t\,f'_{\pi\pi}/m^2_\pi\ , \qquad\ \ f_{\pi K}(t)=f_{\pi K}+(1-t/m
^2_\pi)\,t\,f'_{\pi K}/m^2_\pi\ .\end{equation} In the best of the three fit
variants given in Ref. [5], $M_1=0.773$ GeV, $M_2=1.163$ GeV, $g_\pi=0.848$
GeV, $g'_\pi=0.0479$, $G_\pi=0.848$ GeV, $G'_\pi=-0.0259$ GeV, $f'_{\pi\pi}=0.0
963$, $f_{\pi K}=0.687$, and $f'_{\pi K}=0.0818$. It follows from Eq. (5) that
$g_\pi(t)$ vanishes at $t\approx-0.0728$ GeV$^2$. \footnote{Furthermore, as $-t
$ varies from 0 to 1 GeV$^2$, the functions $g^2_\pi(t)$ and $G^2_\pi(t)$
increase, respectively, by approximately factors of 22000 and 6000. The
appearing enormous rise with $-t$ of the amplitude $a_{\pi\pi}(m,t)$ in Eq. (1)
is compensated by the very rapidly dropped form factor $F(t)=[(\Lambda-m
^2_\pi)/(\Lambda-t)]^4$ with $\Lambda=0.1607$ GeV$^2$ which the authors of Ref.
[5] ascribed to the nucleon vertex (see also Refs. [6,7,10]). The critical
discussion of such an ascription leading to unsolvable difficulties in
different reactions has been given in Ref. [8]. For example, the above form
factor would yield an abnormally sharp drop of the one-pion exchange (OPE)
contribution to the differential cross section of the charge exchange reaction
$pn\to np$. Since $d\sigma^{(OPE)}(np\to pn)/dt\sim|F(t)|^4$, then, in the $-t$
region from 0 to 0.2 GeV$^2\,$ this cross section drops like $\exp(56t)$, which
is comparable only to the fall of the cross sections of diffractive processes 
on complex nuclei.} 
Hence, with increasing $-t$, a dip in the $(\pi^0\pi^0)_S$ mass
spectrum in the $f_0(980)$ region gradually disappears and eventually
turns into a resonancelike enhancement [5]. Here it is worth noting that the
amplitude (2) on the mass shall [$a_{\pi\pi}(m,t=m_\pi^2)$] vanishes at $m=m_0
\approx0.986$ GeV, i.e. just below the $K\bar K$ threshold, due to the
destructive interference between the various contributions, and that the phase
shift of $a_{\pi\pi}(m,t=m_\pi^2)$ goes through 180$^\circ$ at this point in
close agreement with the experimental data [12,13]. Analyzing the model of Ref.
[5] we revealed that, as $-t$ increases, the amplitude (2) also vanishes but 
for different values of $m<2m_K$. The zero ``trajectory"\  of the amplitude (2)
in the plane of the $m$ and $t$ variables is shown in Fig. 1. It is seen that
with increasing $-t$ the amplitude zero shifts, gradually speeding up, from the
region of $m\approx2m_K$ to the lower mass region. For example, as $-t$
increases from 0.09 GeV$^2$ only by 0.026 GeV$^2$, it crosses the wide region
of $m$ from 0.91 to 0.60 GeV.

Thus, we discover at once two striking predictions of the POPE model [5].
First, for each fixed $(\pi^0\pi^0)_S$ invariant mass value $m<2m_K$ (or more
precisely, for each small $m$ bin) the presence of a dip in the $t$
distribution, $dN/dt$, is predicted in the low $-t$ region. For example, in any
interval of $m$ from the region $0.6<m<0.91$ GeV, a dip in $dN/dt$ must be
located near $-t\approx0.1$ GeV$^2$, and, as $m$ increases from 0.91 to 0.986
GeV, it must move to $t=0$. Second, the model predicts that the $m$
distribution of the $\pi^-p\to(\pi^0\pi^0)_S\,n$ reaction events, $dN/dm$, for
$0.6<m<0.9$ GeV must be suppressed in the vicinity of $-t\approx0.1$ GeV$^2$
because in this region of the variables the one-pion exchange amplitude is
close to zero, but, for $m>0.9$ GeV it must sharply increase. Thus, the model
of Ref. [5] describing the GAMS data [1] on the alteration of the $(\pi^0\pi^0)
_S$ mass spectrum in the $f_0(980)$ resonance region for $-t>0.3$ GeV$^2$ can
be unambiguously checked owing to its predictions for the $dN/dt$ and $dN/dm$
distributions for $0<-t<(0.2-0.25)$ GeV$^2$ and $0.6\mbox{\,GeV}<m<2m_K$.
Certainly, to do this much more detailed data are required than those presented
by the GAMS Collaboration. Let us recall that the GAMS data [1] on the reaction
$\pi^-p\to(\pi^0\pi^0)_S\,n$ include single $dN/dm$ distribution in the region
$0.8<m<1.2$ GeV for $0<-t<0.2$ GeV$^2$ (i.e. for the low $-t$ region as a
whole) and, in addition, the $dN/dm$ distributions in the region $0.6<m<1.4$
GeV for five overlapping intervals of $-t$ covering the region $0.3<-t<1$ GeV$
^2$.

In the subsequent versions [6,7,9,10] of the POPE model [5], the $K$ matrix
analysis of the $IJ^{PC}=00^{++}$ waves has been extended to the more wide
regions of $m$ and a larger number of the coupled channels. In Ref. [6], four
resonances coupled to $\pi\pi$, $K\bar K$, $\eta\eta$, and $4\pi$ channels have
been included in the $K$ matrix and the region up to 1.55 GeV has been
analyzed. Five resonances coupled to five channels have been taken into account
in Refs. [7,9,10] and the region of the data description has been extended up
to 1.9 GeV. Certainly the further resonances with masses in the range $1.2-1.9$
GeV [6,7,9,10] exert some influence on the mass region below 1 GeV. However,
with the exception of some details, all essential predictions of the
two-resonance model [5] for $m<1$ GeV remain valid. For example, the most
essential feature of the one-pion exchange amplitude parametrization proposed
in Ref. [5], namely, the passage through zero of the residue of the lowest-mass
resonance with increasing $-t$, takes place in all subsequent variants. The 
mass of the lightest resonance varies with the $K$ matrix parametrization way
from 0.65 to 0.86 GeV [5-7,9,10]. According to the best fit of Ref. [5] the
residue of the resonance vanishes at $-t=0.0728$ GeV$^2$ (this fact has been
already mentioned above),  according to Ref. [6] (solution I) at $-t=0.117$
GeV$^2$, according to Ref. [7] (solution I) at $-t=0.0683$ Gev$^2$, and
according to Ref. [9] at $-t=0.038$ GeV$^2$. Unfortunately, in Ref. [10], the
parameter values needed for the determination of the zero location are absent.

Note that after the publication of our work [8] involving new explanation
of  the GAMS results and criticism of the POPE model [5] the $a_1$ exchange
contribution also appeared in Ref. [9]. However, this contribution was taken
into account in Ref. [9] by a `` purely cosmetic way"\  since, in doing so,
the parametrization of the one-pion exchange amplitude and its dominant role in
the description of the observed alteration phenomenon actually left unchanged.
As the $a_1$ exchange contribution is really small in the low $-t$ region, it 
is reasonable that the predictions of the model [9] for small $-t$ and $m<1$
GeV as a whole turned out to be close to those of the POPE model [5] which were
qualitatively described above. In fact, this claim can be done immediately on
inspection of Figs. 3 and 5 of Ref. [9] showing the predicted $m$ and $t$
distributions of the $\pi^-p\to(\pi^0\pi^0)_S\,n$ events. It is revealing that
in the last publication [10] the authors again do not take into account the $a
_1$ exchange mechanism as previously in Refs. [5-7].

\begin{center}{\bf III. COMPARISON WITH THE BNL DATA}\end{center}

The BNL-E852 Collaboration presented the high-statistics $m$ distributions of
the $\pi^-p\to(\pi^0\pi^0)_S\,n$ reaction events in the region $2m_\pi<m<2.2$
GeV with the 0.04 GeV-wide step in $m$ for nine sequential fine bins in $t$
covering the region $0<-t<0.4$ GeV$^2$ and for a single wide $-t$ interval from
0.4 to 1.5 GeV$^2$ [4]. The BNL data, which we use to check the predictions of 
the POPE model [5], are shown in Figs. 2 and 3. Let us stress that we are not
concerned with the fitting of these data in the framework of the POPE model 
[5]. We just use the model with those values of its parameters which ensure the
best fit to the GAMS data [1] and compare its predictions with the BNL data [4]
both on the $m$ distributions pertaining to the six fine $t$ bins covering the
region $0<-t<0.2$ GeV$^2$ and on the $t$ distributions for six 0.04
GeV-wide intervals in $m$ which we selected as an example from the region
$0.6<m<1.12$ GeV. Similar detailed distributions have not been presented by the
GAMS Collaboration [1,2]. The only parameter the value of which is needed to be
determined once again is the overall normalization constant $C$ in Eq. (1). We
found this parameter by normalizing the theoretical distribution to the total
number of events in the interval $0.6<m<1.2$ GeV for $0.01<-t<0.03$ GeV$^2$.
The data on the distribution $dN/dm$ for this region of the $m$ and $t$
variables are shown in Fig. 2b. Note that among all the isometric intervals of
$t$ the interval $0.01<-t<0.03$ GeV$^2$ contains the maximal number of the
$\pi^-p\to(\pi^0\pi^0)_S\,n$ events in the region $0.6<m<1.2$ GeV. We consider
such a choice of overall normalization to be quite applicable to give a 
descriptive comparison between the experimental and theoretical distributions
in $m$ and $t$.

Figure 2 shows that there is a satisfactory qualitative agreement of the
experimental and theoretical distributions $dN/dm$ in the intervals $0<-t<0.01$
GeV$^2$ and $0.01<-t<0.03$ GeV$^2$. However, with increasing $-t$, the shape of
the theoretical distributions in $m$ sharply changes. Note that this fact is in
line with the expectations given in Sec. II. In addition, it can be seen from 
Fig. 2 that, according to the POPE model [5], the transformation of a dip in 
the $f_0(980)$ region to a resonancelike bump occurs in the $-t$ range from 0.1
to 0.2 GeV$^2$, i.e. too rapidly. As is also clearly seen from Fig. 2, the
experimental distributions $dN/dm$ leave, in fact, similar to each other
throughout the low $-t$ region from 0 to 0.2 GeV$^2$ and all of them have a dip
on the place of the $f_0(980)$ resonance. Let us emphasize again that unlike
the detailed information presented by the BNL-E852
Collaboration for $0<-t<0.2$ GeV$^2$, the GAMS Collaboration has presented for
this $t$ region, containing some 90\% of all $\pi^-p\to(\pi^0\pi^0)_S\,n$
events, a single ``global"\  distribution $dN/dm$, and it is precisely this
rough one that has been fitted successfully by using the POPE model [5].

The BNL data [4] on the $t$ distributions and the corresponding theoretical
predictions are shown in Fig. 3. It is seen that the POPE model [5] predicts 
the presence of a dip in these distributions in the low $-t$ region if $m<2m_K
$. In direct contradiction, no such dip is observed in reality.

It is evident that the character of the POPE model predictions cannot be 
altered if the finite experimental resolutions in $m$ and $t$ are taken into
account in the construction of the theoretical curves. In any case, the 
agreement of the model with the BNL data cannot be improved essentially.

\begin{center}{\bf IV. CONCLUSION}\end{center}

The question whether the observed alteration of the $(\pi^0\pi^0)_S$ mass
spectrum in the reaction $\pi^-p\to(\pi^0\pi^0)_S\,n$ with increasing $-t$ can
be described exclusively in terms of the amplitude with quantum numbers of the
$\pi$ Regge pole in the $t$ channel is absolutely valid and deserves to be
thoroughly considered. Therefore the first attempt to solve this question 
undertaken in Ref. [5] was of great importance. In our opinion, the merit of 
this work is the formulation of the particular one-pion exchange model 
containing some clear predictions which can be easily tested by experiment. The
above analysis shows that these predictions are in rough
contradiction with the detailed BNL data on the $m$ and $t$ distributions of
the $\pi^-p\to(\pi^0\pi^0)_S\,n$ events. However, from our point of view,
it is valuable that the way outlined in Ref. [5] has been completed 
conclusively. Note that the GAMS Collaboration selected the highest statistics
on the reaction $\pi^-p\to\pi^0\pi^0n$ [1,2], that is why the publication of
their $m$ and $t$ distributions of the $(\pi^0\pi^0)_S$ production events for
fine $t$ and $m$ bins for $0<-t<0.2$ GeV$^2$ and $m<1$ GeV is highly desirable.

In accordance with the aforesaid discussion, it is pertinent also to note that
those consequences that were extracted in Refs. [9] and [17] from the analyses
of the experimental data based on the models of Refs. [5-7,9,10] are not 
justified. \vspace*{0.3cm}

The present work was supported in part by the grant INTAS-RFBR IR-97-232.
\vspace*{0.5cm}

\begin{center}{\bf{FIGURE CAPTIONS}}\end{center}\vspace*{0.3cm}

{\bf Fig. 1.} The zero ``trajectory"\  of the amplitude $a_{\pi\pi}(m,t)$ in
the model of Ref. [5] in the plane of the $m$ and $-t$ variables.
\vspace*{0.4cm}

{\bf Fig. 2.} The $(\pi^0\pi^0)_S$ mass spectra, $dN/dm$, in the reaction $\pi
^-p\to\pi^0\pi^0n$ for six sequential intervals of $-t$ shown just in the 
plots. The data are from the BNL-E852 Collaboration [4]. The curves are
constructed by using Eqs. (1)--(3), (5), and (6). The used values of the 
parameters are mentioned in the text.
\vspace*{0.4cm}

{\bf Fig. 3.} The $t$ distributions, $dN/dt$, of the $\pi^-p\to(\pi^0\pi^0)_S
\,n$ reaction events for six intervals of the invariant mass of the $(\pi^0\pi
^0)_S$ system, $m$, shown just in the plots. The data are from the BNL-E852
Collaboration [4]. Here, as well as in Fig. 14 of Ref. [4], the data for the
intervals $0<-t<0.01$ GeV$^2$ and $0.01<-t<0.03$ GeV$^2$ are combined. The 
curves are constructed by using Eqs. (1)--(3), (5), and (6). The used values of
the parameters are mentioned in the text.

\newpage\begin{figure}\centerline{\epsfysize=9in\epsfbox{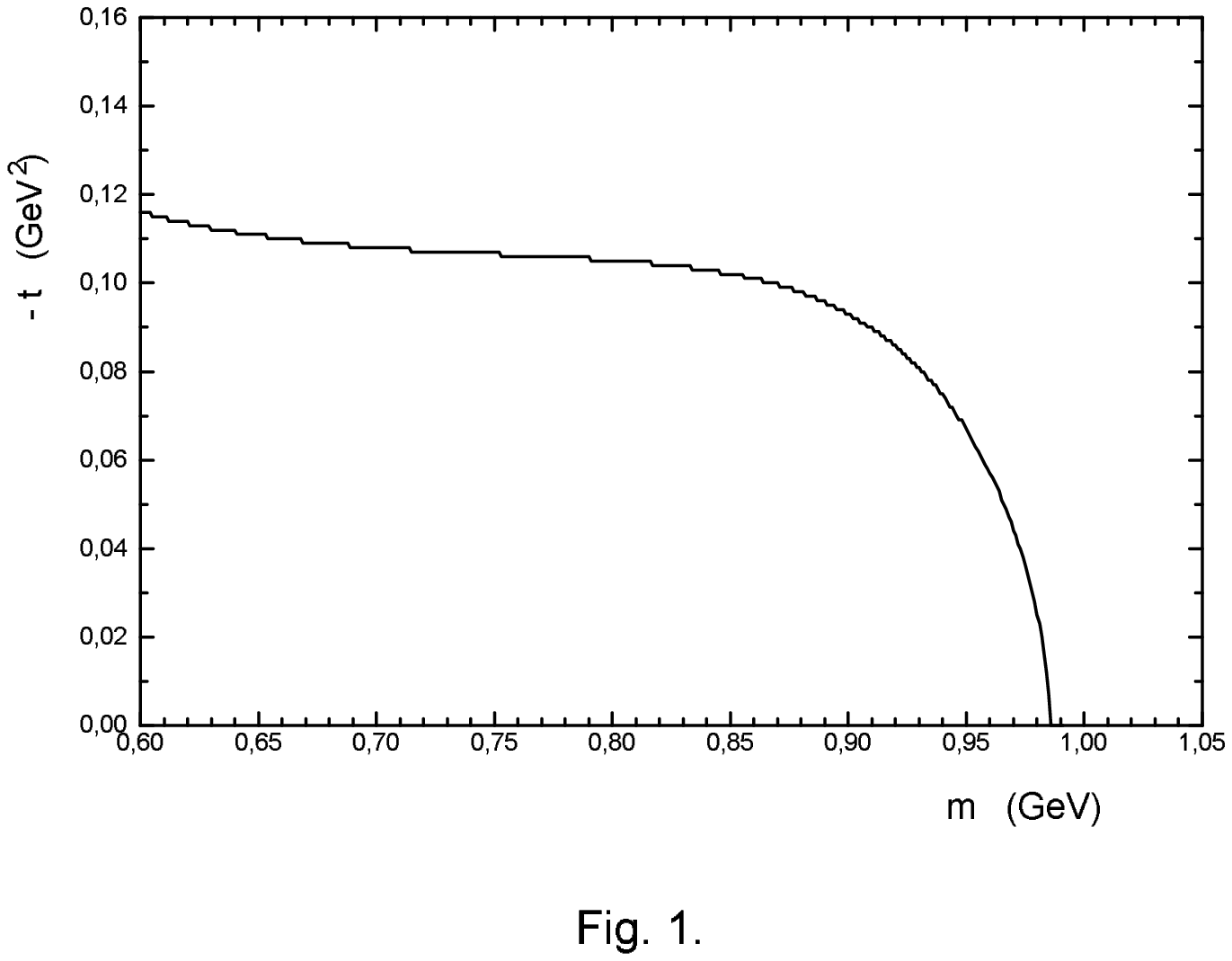}}
\end{figure}

\newpage\begin{figure}\centerline{\epsfxsize=17cm\epsfysize=25cm
\epsfbox{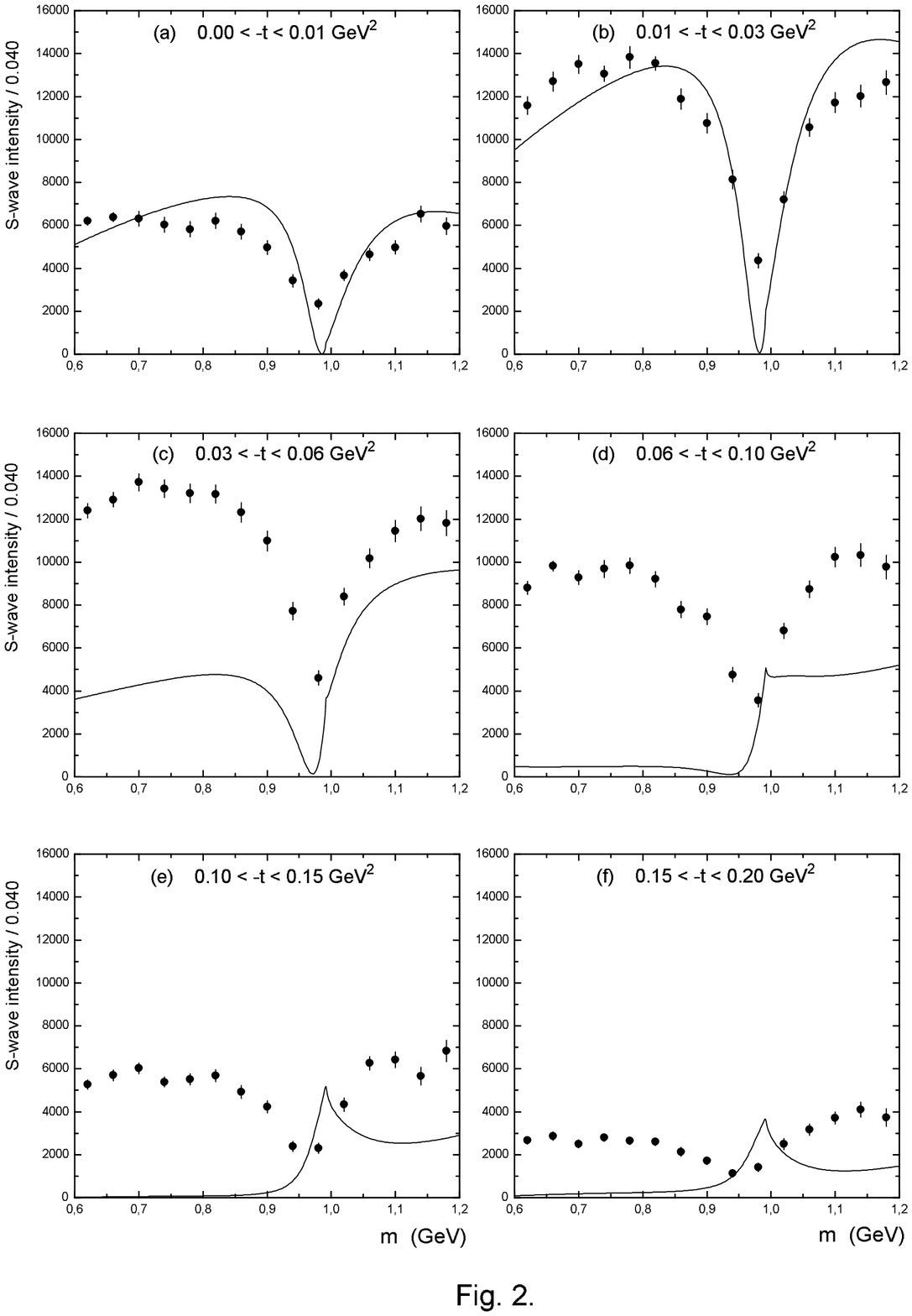}}
\end{figure}

\newpage\begin{figure}\centerline{\epsfxsize=17cm\epsfysize=25cm
\epsfbox{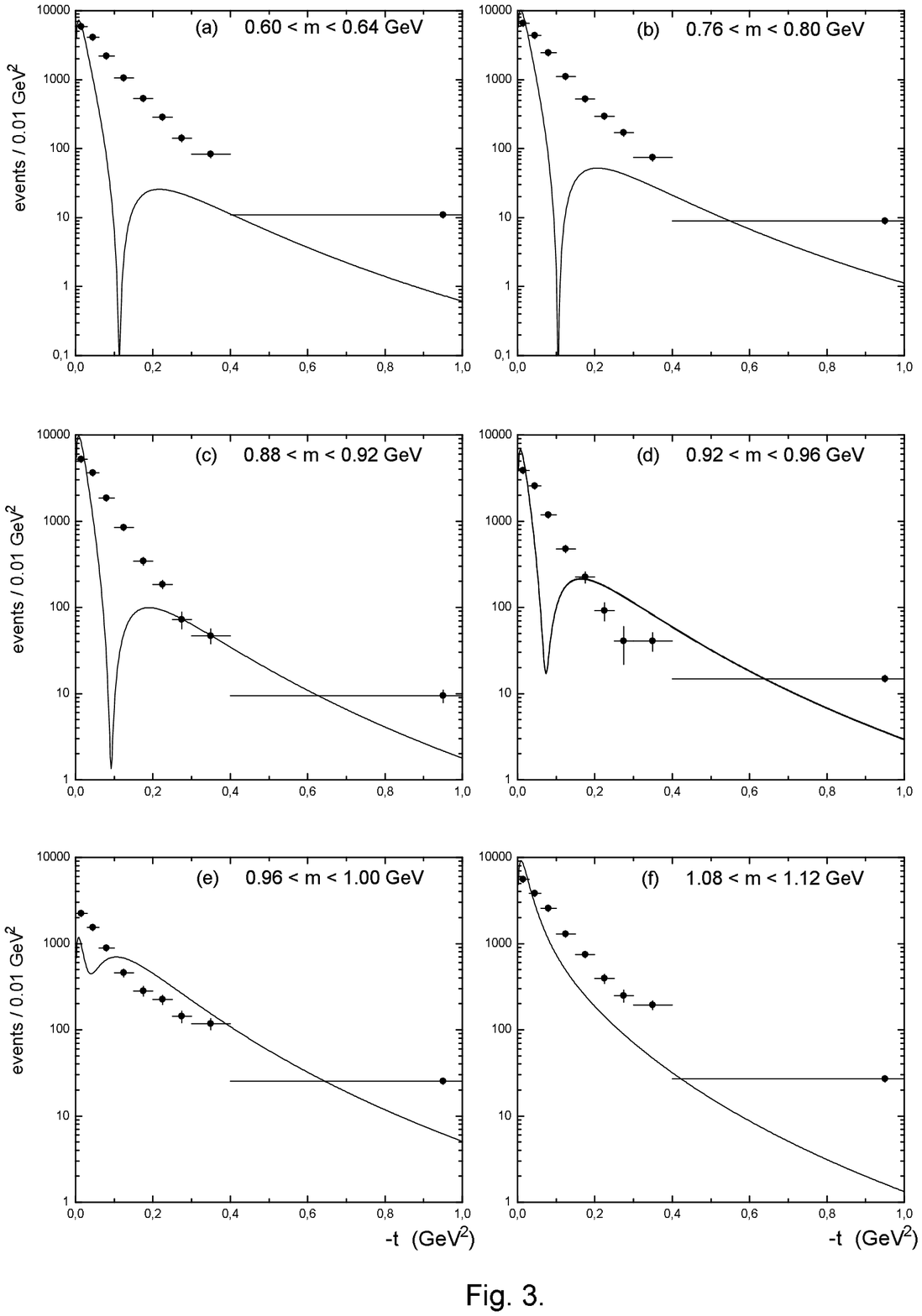}}
\end{figure}

\begin{thebibliography}{99}
\bibitem{} D. Alde {\it et al.,} Z. Phys. C {\bf 66}, 375 (1995).
\bibitem{} Yu.D. Prokoshkin, A.A. Kondashov, S.A. Sadovsky, Physics Doklady
           {\bf 342}, 473 (1995).
\bibitem{} B.B. Brabson, in {\it Proceedings of the 6th International
           Conference on Hadron Spectroscopy}, HADRON '95, Manchester, UK,
           1995, edited by M.C. Birse, G.D. Lafferty, and J.A. McGovern
           (World Scientific, Singapore, 1996), p. 494; A.R. Dzierba, Nucl.
           Phys. {\bf A623}, 142c (1997).
\bibitem{} J. Gunter {\it et al.,} ``A partial wave analysis of the $\pi^0\pi^0
           $ system produced in $\pi^-p$ charge exchange collisions",
           hep-ex/0001038 (submitted to Phys. Rev. D). The detailed results of
           the partial wave analysis are available on the World Wide Web:
           http://dustbunny.physics.indiana.edu/pi0/pi0pwa/.
\bibitem{} V.V. Anisovich {\it et al.,} Phys. Lett. B {\bf 355}, 363 (1995).
\bibitem{} V.V. Anisovich, A.V. Sarantsev, Phys. Lett. B {\bf 382}, 429 (1996).
\bibitem{} V.V. Anisovich, Yu.D. Prokoshkin, and A.V. Sarantsev, Phys. Lett. B
           {\bf 389}, 388 (1996).
\bibitem{} N.N. Achasov and G.N. Shestakov, Phys. Rev. D {\bf 58}, 054011
           (1998) [hep-ph/9802286].
\bibitem{} V.V. Anisovich, D.V. Bugg, A.V. Sarantsev, Phys. Lett. B {\bf 437},
           209 (1998).
\bibitem{} V.V. Anisovich {\it et al.,} Yad. Fiz. {\bf 63}, 1489 (2000).
\bibitem{} P.E. Schlein, Phys. Rev. Lett. {\bf 19}, 1052 (1967);
           E. Malamud and P.E. Schlein, {\it ibid.} {\bf 19}, 1056 (1967).
\bibitem{} B. Hyams {\it et al.,} Nucl. Phys. {\bf B64}, 134 (1973);
           {\bf B100}, 205 (1975).
\bibitem{} P. Estabrooks and A.D. Martin, Nucl. Phys. {\bf B79}, 301 (1974);
           {\bf B95}, 322 (1975).
\bibitem{} W. Hoogland {\it et al.,} Nucl. Phys. {\bf B69}, 266 (1974);
           {\bf B126}, 109 (1977).
\bibitem{} M. Svec, Phys. Rev. D {\bf 53}, 2343 (1996).
\bibitem{} R. Kami\'nski, L. Le\'sniak and K. Rybicki, Z. Phys. C {\bf 74},
           79 (1997).
\bibitem{} V.V. Anisovich, V.A. Nikonov, A.V. Sarantsev, ``Determination of
           hadronic partial widths for scalar-isoscalar resonances $f_0(980)$,
           $f_0(1300)$, $f_0(1500)$, $f_0(1750)$ and the broad state $f_0(1530+
           90-250)$", hep-ph/0102338.ÿ
\end{thebibliography}
\end{document}